\documentclass[aps,prl,superscriptaddress,reprint,longbibliography,amssymb]{revtex4-1}

\usepackage[utf8]{inputenc}
\usepackage[T1]{fontenc}

\usepackage[english]{babel}
\usepackage{letltxmacro}

\DeclareMathAlphabet{\mathbbold}{U}{bbold}{m}{n}

\LetLtxMacro{\ORIGselectlanguage}{\selectlanguage}
\makeatletter
\DeclareRobustCommand{\selectlanguage}[1]{%
  \@ifundefined{alias@\string#1}
    {\ORIGselectlanguage{#1}}
    {\begingroup\edef\x{\endgroup
       \noexpand\ORIGselectlanguage{\@nameuse{alias@#1}}}\x}%
}
\newcommand{\definelanguagealias}[2]{%
  \@namedef{alias@#1}{#2}%
}
\makeatother

\definelanguagealias{en}{english}

\usepackage{physics}
\usepackage{graphicx}
\usepackage[hidelinks]{hyperref}
\usepackage[dvipsnames]{xcolor}
\renewcommand{\imath}{i}

\begin{document}

\title{Orbital angular momentum interference of trapped matter waves}

\author{Filip Kia\l ka}
\affiliation{Faculty of Physics, University of Vienna, Boltzmanngasse 5, A-1090 Vienna, Austria}
\affiliation{Faculty of Physics, University of Duisburg-Essen, Lotharstra\ss e 1, 47048 Duisburg, Germany}
\author{Benjamin A. Stickler}
\affiliation{Quantum Optics and Laser Science, Imperial College London, SW72AZ London, United Kingdom}
\affiliation{Faculty of Physics, University of Duisburg-Essen, Lotharstra\ss e 1, 47048 Duisburg, Germany}
\author{Klaus Hornberger}
\affiliation{Faculty of Physics, University of Duisburg-Essen, Lotharstra\ss e 1, 47048 Duisburg, Germany}

\date{\today}

\begin{abstract}
We introduce a matter wave interference scheme based on the quantization of  orbital angular momentum in a ring trap. It operates without beam splitters, is sensitive to geometric phases induced by external gauge fields, and allows measuring interatomic scattering lengths. {We argue that} orbital angular momentum interferometry offers a {versatile} platform for quantum coherent experiments with cold atoms and Bose-Einstein condensates using state-of-the-art technology.
\end{abstract}

\maketitle

\textit{Introduction---} 
Trapped interference experiments~\cite{WuPRL2005, SackettPRA2006, NakagawaPRL2007, karski2009quantum, SchmiedmayerNC2013, SchmiedmayerS2018, neil2019, MullerS2019} are promising platforms for the next generation of force and acceleration sensors.
Guiding matter waves enables atom interferometers with long interrogation times, while providing considerable freedom for choosing the geometry~\cite{GarrawayPRL2001, bell2016bose, BeugnonPRA2017, KlitzingN2019}.
Toroidal traps are particularly attractive for fundamental quantum experiments~\cite{Stamper-KurnPRL2005, PhillipsPRL2007, CampbellPRL2011, ZurekSR2012, CampbellN2014, CampbellPRX2018} and for precision sensing~\cite{Stamper-KurnPRA2015, TaylorPRL2016, GardinerPRL2018} with ultracold gases or fluids.
The ring geometry implies that the orbital angular momentum of the revolving particles is conserved.
As argued in the following, its fundamental quantization can be exploited to realize trapped interference schemes requiring no beam splitters.

{We note that} the free quantum dynamics in a ring geometry exhibit {\it quantum revivals}. {An} initially well-localized wave packet quickly disperses {along} the ring on a timescale determined by the orbital angular momentum spread. Only after a much longer quantum revival time, which is independent of the initial state, does the localized wave packet {briefly} reappear due to the quantization of orbital angular momentum \cite{RobinettPR2004}. Similar revival effects are encountered in the orientation of revolving molecules \cite{SeidemanPRL1999,IvanovPRL2004,poulsen2004nonadiabatic}, and they have been proposed for  electromagnetic pulse shaping in semiconductors~\cite{BerakdarPRB2006} as well as for macroscopic quantum superposition tests with nanorotors~\cite{HornbergerNJP2018}.

Here, we propose an interference scheme which exploits the brief emergence of a balanced superposition at half the revival time.
By imprinting a relative phase { on the superposition,} one can coherently control at which antipode the wave packet reappears after the full revival time.
The presence of an additional gauge field induces a rotation of the revival determined by the accumulated geometric phase. 
In contrast to many existing proposals for interference in ring traps~\cite{Stamper-KurnPRA2015,KlitzingNJP2016,AhufingerNJP2018}, orbital angular momentum interference does not rely on atomic spin states or collective excitations.
It is thus applicable to all matter-wave experiments with a toroidal geometry, ranging from electrons in solid state quantum rings~\cite{Fomin2014} to nanoparticles in optomechanical traps~\cite{VamivakasRPP2020}.
Here we discuss the special case of optically trapped atomic clouds or Bose-Einstein condensates (BECs), and show that this scheme is sufficiently resilient to {be realizable} with state-of-the-art technology.

\textit{Interference scheme---\label{sec:interference_schemes}} In order to explain the  interference scheme we first consider the idealized case of  a point particle of mass $m$ confined to a circle of radius $R$. Its Hamiltonian reads  $\mathsf{H} = \mathsf{L}_z^2 / 2 m R^2$. Since the eigenvalues of the orbital angular momentum operator  $\mathsf{L}_z$ are integer multiples of $\hbar$, with eigenstates $\ket{\ell}$, the time evolution operator $\mathsf{U}_0(t) = \sum_{\ell \in \mathbb{Z}} \exp(-\imath \hbar t \ell ^2 / 2 m R^2) \ketbra{\ell}{\ell}$ is unity for all even multiples of the revival time
\begin{equation} \label{eq:revival-time}
	T_\text{rev} = \frac{2 \pi m R^2}{\hbar}.
\end{equation}

A straightforward calculation shows that the  evolution for the revival time performs a $\pi$ rotation, ${\sf U}_0(n T_{\rm rev}) = \exp (i n \pi {\sf L}_z /\hbar)$, with $n \in \mathbb{N}_0$.
In a similar fashion, free evolution for $T_{\rm rev}/2$ acts as a beam splitter, preparing a balanced superposition of the initial state and its $\pi$-rotated version~\cite{SorokinEJP2003,IvanovPRL2004},
\begin{equation} \label{eq:beamsplitting}
	\mathsf{U}_0 \left(\frac{T_\text{rev}}{2}\right)= \frac{\mathrm{e}^{- \imath \pi / 4}}{\sqrt{2}} \left(\mathbbold{1} + \imath \mathrm{e}^{\imath \pi \mathsf{L}_z / \hbar}\right),
\end{equation}
where $\mathbbold{1}$ is the unity operator.

An initially tightly confined wave packet thus first disperses on a short timescale  determined by its initial angular momentum uncertainty. The state then remains delocalized over the ring for most of time, showing fractional revivals such as Eq.~\eqref{eq:beamsplitting} at fractions of the revival time. The lifetime of these fractional and full revivals is  determined by the initial dispersion time, and is thus  typically orders of magnitude smaller than the revival time itself.

\begin{figure}[bt]
	\centering
	\includegraphics[width=0.4\textwidth]{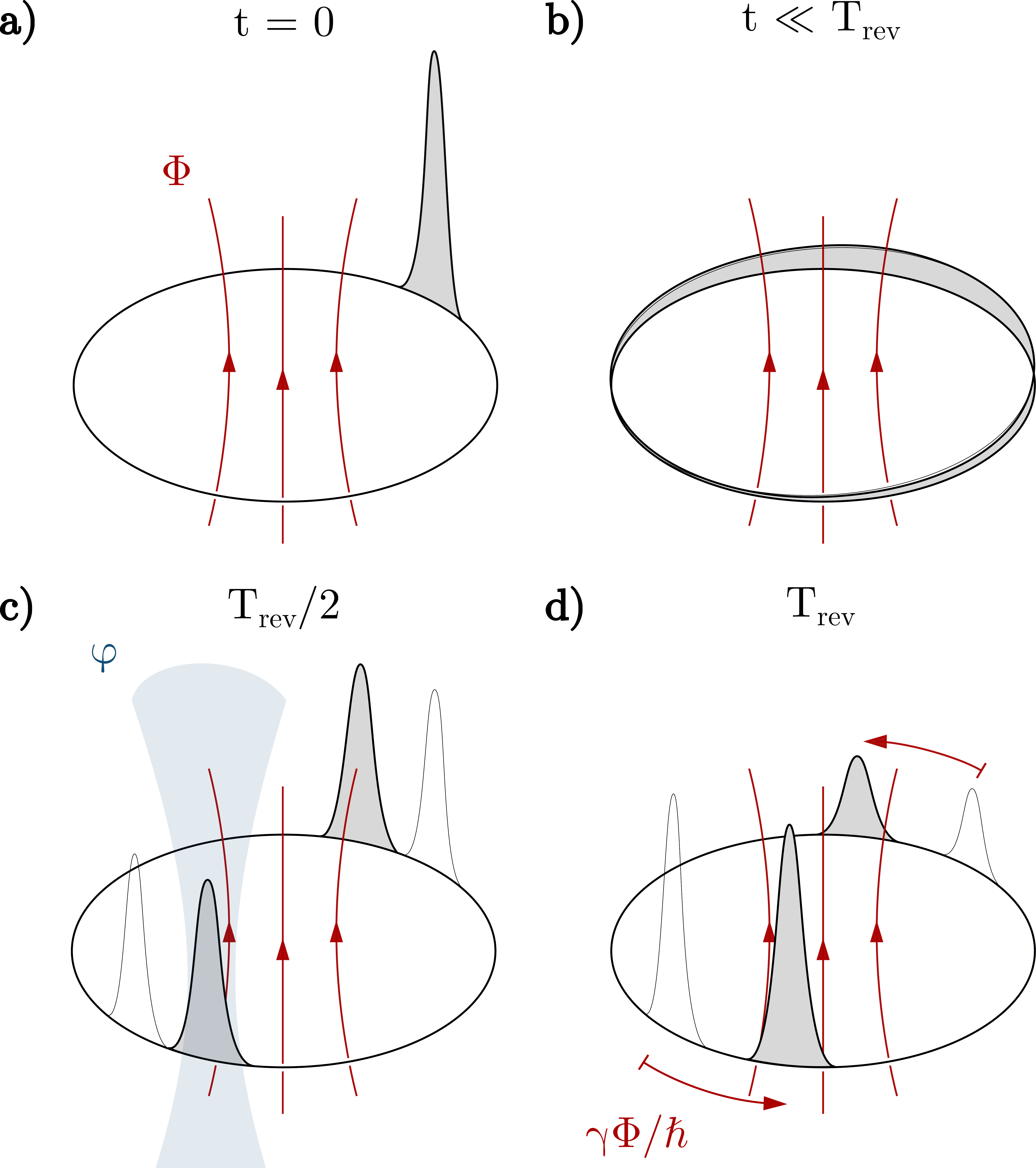}
	\caption{
		{ Schematic illustration of the orbital angular momentum interference effect.  {A localized wave packet (a) quickly disperses (b), before  reappearing (c) at $T_\text{rev} / 2$ for a short period of time in a balanced superposition of the original and mirrored locations; see \eqref{eq:revival-time} and \eqref{eq:beamsplitting}.	Applying a relative phase $\varphi$ between the two superposition components controls  final population imbalance at the antipodes (d) after further time evolution for $T_\text{rev} / 2$.
		If a gauge field is turned on quickly after releasing the wave packet, the interference pattern (black) is rotated with respect to the field-free case (gray) by an angle $\gamma \Phi/\hbar$, equal to the Aharonov-Bohm phase [see Eq. \eqref{eq:U-AB}].}}
	}
	\label{fig:schemes}
\end{figure}

The dynamical beam splitting described by (\ref{eq:beamsplitting}) is exploited by the following interference scheme; see Fig.~\ref{fig:schemes}(a): The particle is initially prepared in a well-localized state $\ket{\psi_0}$. After dispersing on a short timescale, the localized state reappears at half of the revival time { in a}  balanced superposition $(\ket{\psi_0} + i \ket{\psi_\pi})/\sqrt{2}$, with the $\pi$-rotated initial state $\ket{\psi_\pi} = \exp(i \pi {\sf L}_z/\hbar)\ket{\psi_0}$. { Then} a relative phase $\varphi$ is induced between the two wave packets, for instance gravitationally by tilting the ring, optically via laser illumination, or in the case of an atomic cloud via magnetic control of the scattering length. After imprinting  the phase, the state evolves freely for another $T_{\rm rev}/2$, yielding the final state $\ket{\psi_f} = \cos(\varphi/2)\ket{\psi_\pi} + i \sin(\varphi/2)\ket{\psi_0}$. The final position of the particle is thus determined interferometrically.

{\it Gauge fields and external potentials---} The interference effect  depends  sensitively on the interaction with external gauge fields.
If the field ${\bf A}({\bf r})$ is minimally coupled to the canonical angular momentum ${\sf L}_z$, the kinetic angular momentum is ${\sf L}_z - \gamma R A(\hat \alpha)$.
Here $\gamma$ is the gauge coupling and $A(\alpha) = {\bf A}(R {\bf e}_\rho(\alpha)) \cdot {\bf e}_\alpha (\alpha)$ is the azimuthal component of the gauge field evaluated at the angular position~$\alpha$.

The presence of ${\bf A}({\bf r})$ implies a gauge-invariant flux $\Phi = R \oint d \alpha A(\alpha)$ piercing the ring interferometer and thus modifying the free time evolution of the matter wave.
The unitary time evolution operator becomes
\begin{equation} \label{eq:U-AB}
	\mathsf{U}_\Phi \left(t\right) =  \mathsf{V} \exp(\imath \frac{\gamma \Phi}{\hbar} \frac{t}{T_{\rm rev}} \frac{\mathsf{L}_z}{\hbar}) \mathsf{U}_0 (t) \mathsf{V}^\dagger,
\end{equation}
where $\mathsf{V} = \exp(- \imath \gamma \Phi \hat{\alpha} / 2 \pi \hbar + \imath \gamma R / \hbar \int_0^{\hat{\alpha}} d\alpha' A(\alpha'))$ can always be set to unity by choosing an appropriate gauge (symmetric gauge in the case of a constant field).
Thus, a finite flux {induces} a rotation of the recurred wave packet by the Aharanov-Bohm-type phase $\gamma \Phi/\hbar$. 

For example, if the particles are electrically charged, $\gamma = q$, a magnetic flux $\Phi$ through the ring will shift the energy levels \cite{PetroffPRL2000,SorokinEJP2003} causing the  wave packet to rotate.
In a similar fashion, the Aharonov-Casher phase~\cite{CasherPRL1984} can be measured 
if a magnetic dipole ${\bf m} = m_0 {\bf e}_z$ evolves in presence of  the electrostatic field ${\bf E}(R{\bf e}_\rho) = E_0 {\bf e}_\rho$ produced by a line charge. In this case one has $\gamma {\bf A} = {\bf m} \times {\bf E}/c^2$, implying $\gamma \Phi = 2 \pi R E_0 m_0 / c^2$. Likewise, geometric phases can result for a permanent or induced electric dipole ${\bf p}$ in a magnetostatic field ${\bf B}$, so that $\gamma{\bf A} = \vb{p} \times \vb{B}$ \cite{wei1995}, or for a massive particle in a noninertial frame rotating with angular frequency $\boldsymbol{\omega}$ around the trap center, so that $\gamma {\bf A} = m R^2 {\boldsymbol \omega}$.

The presence of a weak external potential $V(\alpha) = V_0 \cos(\alpha - \alpha_0)$, such as that arising from a constant tilt of the ring, leads to phase dispersion.
To leading order in $V_0$, the energies are shifted by
\begin{equation}
    \Delta E_\ell^{\rm (pot)} \approx \frac{m R^2 V_0^2}{4 \hbar^2} \left (\ell^2 - \frac{1}{4} \right )^{-1}.
\end{equation}
Since this is not proportional to $\ell^2$, a conservative torque affects the shape of the recurring wave packet. This is in contrast to gauge fields, which only shift the position { of the revival}.

{\it Revivals in 3D torus traps---} The evolution of a particle in a real-world (three-dimensional) torus trap differs from the idealized situation described so far. The dynamics transverse to the ring tangent affect the angular dynamics even if the transverse motion remains in its ground state, since { the centrifugal force distorts the level spacing}. Shape imperfections and excitations of the transverse degrees of freedom can further affect the interference.
We will show next that the proposed orbital angular momentum interference protocol is { nevertheless} surprisingly robust and remains feasible for realistic trap geometries.

To study { the dynamics in a real-world torus trap}, we expand the full 3D Hamiltonian of a particle in a torus trap and consider leading-order corrections in the transverse size of the wave packet. For this sake, we use { a} Frenet-Serret coordinate system $(s,u,v)$ with  arc length $s $ and two transverse coordinates $u,v$. Thus, the position vector is $\vb{r} = \vb{R}(s) + u \vb{n}(s) + v \vb{b}(s)$, where $\vb{R}(s)$ traces the center {line} of the torus trap, while $\vb{n}(s) = {\bf R}''(s)/\kappa$ and $\vb{b}(s) = {\bf R}'(s) \times {\bf n}(s)$ span the transverse plane at each position~\cite{PavloffPRA2001,StringariNJP2006}. Here, $\kappa = | {\bf R}''(s) |$ is the curvature, where prime denotes derivative with respect to $s$.

Since the new coordinate system $(s,u,v)$ is curved, coordinate-space normalization of the wave function includes the root {of the} metric determinant {  (Jacobian) $h$}.
Expressing the latter as { $h = 1 - \kappa u$} and assuming that the trapping potential is separable in the transverse direction yields the Hamiltonian~\cite{PavloffPRA2001,StringariNJP2006}
\begin{align} \label{eq:H}
	\mathsf{H}_s = {}& - \frac{\hbar^2}{2 m} \left[\partial_s \frac{\partial_s}{h^2} + \partial_u^2 + \partial_v^2 + \frac{\kappa^2}{4 h^2} + \frac{5 (h')^2}{4 h^4} - \frac{h''}{2 h^3}\right] \nonumber\\
	{}& + V_u(u) + V_v(v),
\end{align}
which acts on the rescaled wave function { $\chi = \sqrt{h} \psi$}.

If the radially confining potential is harmonic with frequency $\omega_\perp$ and assuming that centrifugal distortions and small deviations from the ideal circular trap can be described by expanding the Hamiltonian to first order in the small quantities $\kappa \sigma_u$, $\kappa' \sigma_u /\kappa$, and $\kappa'' \sigma_u / \kappa^2$ (with $\sigma_u=\sqrt{\hbar /m\omega_\bot}$ the width of the transverse ground state),
\begin{align}
	\mathsf{H}_s \approx& - \frac{\hbar^2}{2 m} \bigg[\left(1 + 2 \kappa u \right) \left(\partial_s^2 + \frac{\kappa^2}{4}\right) + 2 \kappa' u \left(1 + 3 \kappa u \right) \partial_s \nonumber \\
	&{} + \frac{\kappa^{\prime\prime}u}{2} + \partial_u^2 + \partial_v^2 \bigg] + \frac{m \omega_\perp^2}{2}u^2 + V_v(v). \label{eq:H-first}
\end{align}

\textit{Centrifugal energy corrections---\label{sec:imperfections}} For { an ideal} torus where $\kappa = 1/R$ the stationary Schrödinger equation becomes separable. It admits solutions of the form
\begin{equation}
	\chi_{\ell k n} (s, u ,v) = \frac{1}{\sqrt{2 \pi R}} \mathrm{e}^{\imath \ell s / R} \xi_{\ell k}(u) \Psi_n(v),
\end{equation}
with eigenenergies $E_{\ell k n} = \hbar^2 \ell^2/2 m R^2 + E_{\ell k}^{(u)} + E_n^{(v)}$ where $k,n \in \mathbb{N}_0$. Here, $\Psi_n(v)$ are normalized eigenstates of the harmonic motion out of the ring plane, whose eigenenergies $E_n^{(v)}$ are independent of $\ell$ and thus do not affect the revival structure of the matter wave.

The radially confining harmonic potential in the Schr\"{o}dinger equation for $\xi_{\ell k}(u)$ is centrifugally shifted by $u_\ell = \hbar^2\left(\ell^2 - 1/4\right)/m^2 \omega_\perp^2 R^3$,
\begin{align}  \label{eq:radial-schreq}
	\left [ - \frac{\hbar^2}{2 m} \partial_u^2 + \frac{m \omega_\perp^2}{2}  \left(u^2 + 2 u u_\ell \right ) \right ]\xi_{\ell k}(u) = E^{(u)}_{\ell k} \xi_{\ell k}(u).
\end{align}
Thus, the eigenergies
\begin{equation} \label{eq:centrifugal-shift}
	E_{\ell k}^{(u)} = \hbar \omega_\bot \left ( k + \frac{1}{2} \right ) - \frac{\hbar^4}{2 m^3 \omega_\perp^2 R^6} \left(\ell^2 - \frac{1}{4}\right)^2,
\end{equation}
are lowered due to the centrifugal barrier.

The $\ell$ dependence in the eigenenergies \eqref{eq:centrifugal-shift} can shift and diminish the revival. Specifically, the $\ell^2$ term in Eq.~\eqref{eq:centrifugal-shift} { delays} the revival  without affecting its visibility, while the $\ell^4$ correction decreases the fidelity of the revival and may further modify the revival time. The optimal recurrence time can be determined numerically from this equation.

{\it Shape imperfections ---} In practice, deviations from the perfect circular shape of the torus trap are the most important source of imperfections for optical traps. In particular, residual astigmatism in the focusing optics may introduce a finite ellipticity to the trap, which can be quantified with the help of \eqref{eq:H-first}. 

We replace the arc length with the eccentric anomaly $\beta \in [- \pi, \pi)$ used for the standard parametrization of the ellipse. Thus, $\partial_s =  h_{\varepsilon}^{-1}(\beta) \partial_\beta/R$, where $R$ and $\varepsilon$ are the semimajor axis and the eccentricity and $h_{\varepsilon}(\beta) = \sqrt{1 - \varepsilon^2 \cos^2 \beta}$ is the Jacobi determinant of the ellipse. In lowest order of $\varepsilon$, the Hamiltonian reads as ${\sf H}_\beta = h_{\varepsilon}^{1/2} {\sf H}_s h_{\varepsilon}^{-1/2} \approx {\sf H}_\beta^{(0)} + \varepsilon^2 {\sf H}_\beta^{(\varepsilon)}$, where ${\sf H}_\beta^{(0)}$ describes the motion on the circle and
\begin{align} \label{eq:eccentricity}
	\mathsf{H}_\beta^{(\varepsilon)} ={}& -\frac{\hbar^2}{4mR^2} \left[1 + \frac{3u}{R} +\left( 1 + \frac{5 u}{R} \right) \cos(2 \beta)\right] \partial_\beta^2 \nonumber\\
	&{} +\frac{\hbar^2}{2 m R^2} \left(1 + \frac{5u}{R} + \frac{9 u^2}{R^2} \right) \sin (2 \beta) \partial_\beta \nonumber\\
	&{} -\frac{\hbar^2}{16 m R^2} \left[1 + \frac{3 u}{R} - \left(1 + \frac{11u}{R} \right) \cos (2\beta) \right],
\end{align}
This implies that the eccentricity-induced energy shift reads in first order perturbation theory
\begin{equation} \label{eq:ellipticity-shift}
	\Delta E_\ell^{(\varepsilon)} = \frac{\hbar^2 \varepsilon^2}{8 \pi m R^2} \left(1 + \frac{3 u_\ell}{R}\right) \left(\ell^2 - \frac{1}{4}\right).
\end{equation}
Here we expressed the position expectation value of the radial state by the centrifugal shift of the harmonic potential \eqref{eq:radial-schreq}, $\langle u \rangle = - u_\ell$. The { first-order} influence of a finite eccentricity is thus to decrease the revival time, while further diminishing the revival due to the $\ell$ dependence of the radial potential minimum $u_\ell$.

\begin{figure*}[tb]
	\centering
	\includegraphics[width=\textwidth]{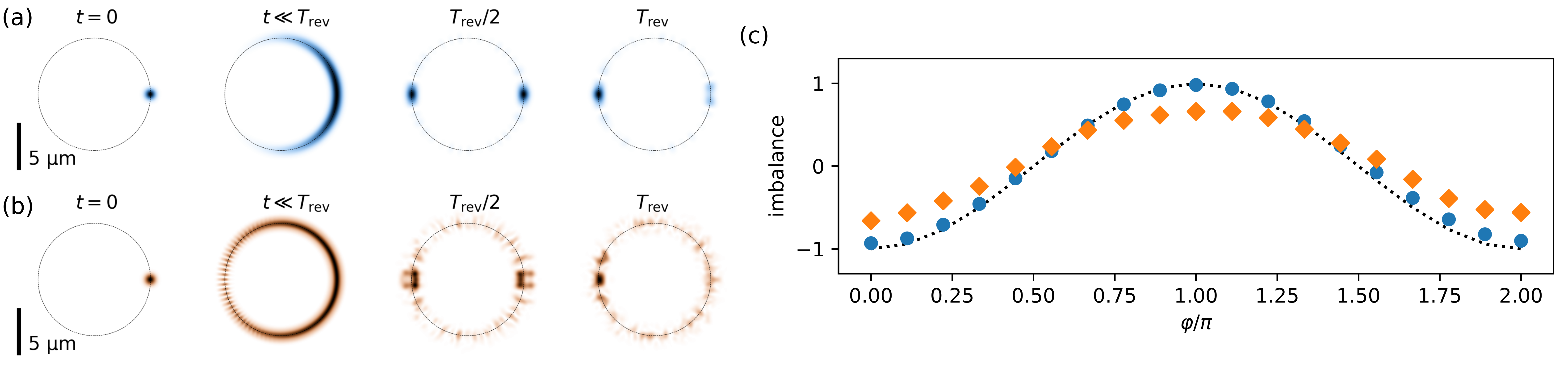}
	\caption{
		Mean-field simulation of the interference scheme shown in Fig.~\ref{fig:schemes}(a) realized with a BEC of $^{39}\mathrm{K}$ in an optical trap.
		(a) Snapshots of the time evolution  for a noninteracting condensate: initial particle density, dispersion, recurrent superposition at half of the revival time, and final interferometrically controlled revival with $\varphi = \pi/3$.
		The external phase of $\exp(i \varphi\cos^2 \alpha)$ is applied on the left part of the ring at $T_\text{rev}/2$.
		The revival time $T_\text{rev} \approx 135.8 \, \text{ms}$ is found by maximizing the overlap between the initial and final states for $\varphi = 0$.
		(b) As in (a) but with interatomic interactions characterized  by the scattering length of one Bohr radius for a BEC of $N = 2 \times 10^4$ atoms.
		As a result of the interactions the revival time { changes} to $T_\text{rev} \approx 136.2 \, \text{ms}$.
		(c) Interference signal as a function of external phase $\varphi$ in the noninteracting [as in (a), circles] and interacting [as in (b), diamonds] cases, as compared to the ideal situation (dotted line).
		The population imbalance is defined as $(N_\text{R} - N_\text{L}) / (N_\text{R} + N_\text{L})$, where $N_\text{R}$, $N_\text{L}$ are the numbers of atoms on the right and left sides of the ring, weighted with $\cos^2\alpha$.
		}
	\label{fig:movie}
\end{figure*}

\textit{Implementation with BECs ---\label{sec:BEC}}
We are now in a position to argue that the orbital angular momentum interference scheme can be realistically carried out with weakly interacting BECs in an optical torus trap.
For concreteness, we consider a condensate of $^{39}\mathrm{K}$ in a trap formed by two coaxial Gaussian beams, one repulsive and one attractive, intersected with an attractive light sheet, as in Ref.~\cite{Stamper-KurnPRA2015}. 
The wavelengths of the red- and  blue-detuned laser beams are assumed to be  $830$ and $532 \, \text{nm}$, respectively, with powers of $2$ and $2.5 \, \text{mW}$ as well as waists of $13$ and $5.5 \, \mu\text{m}$.
The light sheet with {the same} wavelength as the red-detuned laser has a power of $10 \, \text{mW}$ and waists of $5$ and $200 \, \mu\text{m}$, so that the trap radius is $R \approx 5.9 \, \mu\text{m}$ and the transverse confining frequency { $\omega_\bot \approx 6.4 \: \text{kHz}$}. The necessary coherence time of $T_\text{rev} \approx 135 \, \text{ms}$ is experimentally within reach~\cite{MullerS2019}.

Figure \ref{fig:movie} shows the simulated dynamics of the orbital angular momentum interference protocol for (a) a noninteracting and (b) a weakly interacting BEC of $N = 2\times 10^4$ $^{39}$K atoms.
We assume in both cases that the Feshbach resonances of $^{39}$K~\cite{TiesingaRMP2010} are used to make the interactions (a) negligibly small or (b) equivalent to a scattering length of one Bohr radius.
The tightly confined initial wave packet, loaded from three-dimensional harmonic trap of frequency $\omega_\bot$, quickly disperses around the torus.
It then reappears in a superposition after approximately $65$\,ms.
The presence of interactions diminishes the revival signal.
However, even at a realistic transverse confinement and interaction strength, the effect is still clearly visible in the population imbalance displayed in panel (c).
The latter shows that the interference visibility exhibits almost the ideal dependence on the imprinted phase. The numerical calculations are based on the Trotter-Suzuki expansion~\cite{Bederian2011,CucchiettiCPC2013,CalderaroCPC2015}.

For this setup, the centrifugal energy shift \eqref{eq:centrifugal-shift} amounts to a few percent of the rotational energy for the highest-populated $\ell$ eigenstates ($\ell \simeq 25$). The corresponding correction to the revival time is at a permille level, but given the quick dispersion time, exact timing {on the scale of a few microseconds} is required to imprint the phase and to observe the revival. In a similar fashion, the corrections of the revival time due to interactions must be accounted for, as has been done numerically in Fig.\,2(b).

The relative phase $\varphi$ can be imprinted, e.g., optically, via tilting of the apparatus, or via induced interatomic interactions. For example, if the trap is briefly tilted at $T_{\rm rev}/2$ the gravitational potential yields the phase $\varphi_g \approx 2 m g R t_\text{d} \sin\theta / \hbar$, where $\theta$ is the tilt angle and $t_\text{d}$ is the revival lifetime.
The latter is the dispersion timescale $t_\text{d} \approx 1/\omega_\perp$ of the initial wave packet of width $\sqrt{\hbar / \omega_\perp m}$. For the above example, this requires tilting with a precision of hundreds of microradians.

Likewise, if the magnetic field on one side of the ring is detuned from the zero crossing of the Feshbach resonance, the matter wave acquires a relative phase
{$\varphi_a \approx 4 \pi \hbar \, a \, n_{\rm BEC}\, t_{\rm d} / m$}, where $ a$ is the induced  scattering length and $n_{\rm BEC}$ is the particle density in the initial state. With this one can measure the scattering length with precision $\Delta a \approx 0.2 a_0$ { (with $a_0$ the Bohr radius)},
on par with state-of-the-art time-of-flight~\cite{SimoniNJP2007} and spectroscopic~\cite{LisdatPRA2008} measurements for $^{39}\mathrm{K}$.

\textit{Conclusions ---\label{sec:summary_and_outlook}}
We introduced orbital angular momentum interference as an attractive platform for trapped matter-wave interferometry in toroidal geometries.
Since the proposed scheme relies on the universal property of orbital momentum quantization, realizations with many different systems can be readily envisioned, e.g., single atoms or BECs in optical traps, ions in electric traps, electrons in solid state quantum rings, as well as molecules and nanoparticles in optical or electrical traps.
For the case of a BEC in an optical trap, we have shown that the protocol is feasible with present-day technology.

The interference effect is sensitive to the presence of gauge fields. In the presence of a magnetic field flux $\Phi$, for instance, the revival of particles with charge $q$ will be displaced by the angle $q \Phi / \hbar$. Assuming that displacements on the size of the initial wave packet can be angularly resolved, fields below $10^{-7} \, \mathrm{T}$ level can be detected with the setup described above.

{\it Acknowledgments ---} We thank Markus Arndt, Thorsten Schumm, and Philipp Haslinger for helpful discussions. F.K. acknowledges support by the Austrian Science Fund (FWF) Project No. W1210-N25, B.A.S. acknowledges funding from the European Union’s Horizon 2020 research and innovation programme under the Marie Skłodowska-Curie grant agreement No. 841040.

% \bibliography{references,bibliography}

\newcommand{\noopsort}[1]{}

\end{document}